%% file: 2006_gwdaw11.tex
\documentclass{iopart}
\usepackage[dvips]{graphicx}
\usepackage{amssymb}
\usepackage{wasysym}

\usepackage{iopams}

\newcommand{\eqref}[1]{(\ref{#1})}
\newcommand{\abs}[1]{\left|#1\right|}
\newcommand{\text}[1]{\hbox{\scriptsize\rm #1}}
\newcommand{\un}[1]{\mathrm{\,#1}}
\newcommand{\mc}[1]{\mathcal{#1}}

\newcommand{\sinc}{\mathop{\rm sinc}\nolimits}

\newcommand{\dcc}{LIGO-P070028-03-Z}
\newcommand{\nhat}{\hat{n}}
\newcommand{\shat}{\hat{s}}
\begin{document}
\title[Virgo/LSC SB search] {Prospects for Stochastic Background
  Searches Using Virgo and LSC Interferometers}
\author{Giancarlo~Cella$^1$, Carlo~Nicola~Colacino$^2$, Elena~Cuoco$^3$,
  Angela~Di~Virgilio$^1$, Tania~Regimbau$^4$, Emma~L~Robinson$^2$ and
  John~T~Whelan$^5$
  (for the LSC-Virgo working group on stochastic backgrounds)}
\address{$^1$ Istituto Nazionale di Fisica Nucleare sez.\ Pisa, 56100 Pisa, Italy}
\address{$^2$ University of Birmingham, Edgbaston, Birmingham, B15 2TT, UK}
\address{$^3$ European Gravitational Observatory, 56021 Cascina (PI), Italy}
\address{$^4$ Dpt.\ ARTEMIS, Observatoire de la C\^ote d'Azur, BP 429
06304 Nice, France}
\address{$^5$ Max-Planck-Institut f\"{u}r Gravitationsphysik
  (Albert-Einstein-Institut), D-14476 Potsdam, Germany}
\begin{abstract}
  We consider the question of cross-correlation measurements using
  Virgo and the LSC Interferometers (LIGO Livingston, LIGO Hanford,
  and GEO600) to search for a stochastic gravitational-wave
  background.  We find that inclusion of Virgo into the network will
  substantially improve the sensitivity to correlations above 200\,Hz
  if all detectors are operating at their design sensitivity.  This is
  illustrated using a simulated isotropic stochastic background
  signal, generated with an astrophysically-motivated spectral shape,
  injected into 24 hours of simulated noise for the LIGO and Virgo
  interferometers.
\end{abstract}
\ead{john.whelan@aei.mpg.de}


\section{Introduction}
\label{s:intro}

There are four kilometre-scale interferometric gravitational-wave
(GW) detectors currently in operation: the 4\,km and 2\,km
interferometers (IFOs) at the LIGO Hanford Observatory (LHO) (known
respectively as H1 and H2), the 4\,km IFO at the LIGO Livingston
Observatory (LLO) (known as L1), and the 3\,km Virgo IFO (known as
V1).  The LLO and LHO IFOs are currently conducting operations at
their design sensitivity, and operate together with the 600\,m GEO600
IFO (known as G1) under the auspices of the LIGO Scientific
Collaboration (LSC).  Virgo's first full science run commences in May
2007.

One of the signals targeted by ground-based GW IFOs is a stochastic GW
background (SGWB), which can be either of cosmological or
astrophysical origin, in the latter case being produced by a
superposition of unresolved sources.  The standard technique to search
for a SGWB looks for correlations in the outputs of multiple
detectors.  We describe in this paper how the inclusion of correlation
measurements involving Virgo could improve the sensitivity of the
current LLO-LHO network.

\section{All-sky sensitivity at design}

\subsection{Observing Geometry}

\begin{figure}
  \begin{center}
     \includegraphics[width=3in,angle=0]{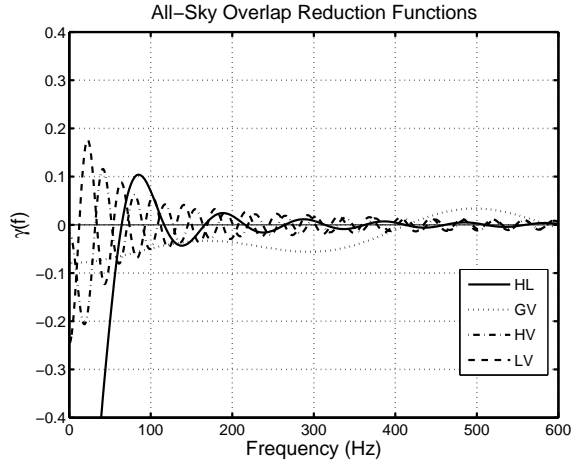}
  \end{center}  
  \caption{
    The overlap reduction functions for pairs of detector sites.
    Note that the ORF for the two LIGO sites goes off the scale of
    this plot at 40\,Hz, which is the ``seismic wall'' below which
    LIGO data are too noisy to be of any use.  The proximity and
    alignment of LLO and LHO makes HL the most favourable pair of
    detector sites for observations below 150\,Hz or so.  However,
    the proximity of the Virgo and GEO600 sites means the GV ORF is
    substantial out to higher frequencies, overcoming the
    low-frequency suppression due to their poor alignment.  On the
    other hand, as shown in Table~\protect\ref{tab:overlap}
    the HV and LV ORFs, while they oscillate
    rapidly with increasing frequency, do not decay as precipitously
    as the HL ORF, making them more favourable
    than HL (but less than GV)
    for $f\gtrsim200\un{Hz}$.
  }
  \label{fig:overlap}
\end{figure}

The effect of a SGWB is to generate correlations in the outputs of a
pair of GW detectors, which can be described for an isotropic
background in the Fourier domain by
\begin{equation}
  \label{eq:h1h2corr}
  \langle \widetilde{h}_1^*(f)\,\widetilde{h}_2(f')\rangle
  = \frac{1}{2}\delta(f-f')\,\gamma_{12}(f)\,S_{\text{gw}}(f)
  \ .
\end{equation}
\begin{table}
  \centering
  \input{gammaDC}
  \caption{Limiting behaviour of overlap reduction
    functions of detector pairs.
    H refers to either of the IFOs at the LHO site, L to LLO, V to Virgo,
    and G to GEO600.
    At $f=0$, the ORF is determined by the alignment of the detectors.
    The reduced inverse light travel time $(2\pi T_{12})^{-1}$ gives
    a characteristic frequency for the onset of ``high-frequency''
    behaviour, which includes a sinc function of the ratio of $f$ to
    that frequency.  However, the limiting form in \eqref{e:gammalim}
    also includes a geometric projection factor, leading to an overall
    envelope $\gamma^{\text{env}}_{12}$ which is shown in the third row.
    In particular, while the light-travel time $T_{HL}$ is less than
    $T_{HV}$ or $T_{LV}$, the projection factor more than makes up
    for this, which makes the mean amplitudes of $\gamma_{\text{HV}}(f)$
    and $\gamma_{\text{LV}}(f)$
    at high frequencies larger than that for $\gamma_{\text{HL}}(f)$.}
  \label{tab:overlap}
\end{table}
The raw correlation depends on the (one-sided) power spectral density
$S_{\text{gw}}(f)$ the SGWB would generate in an IFO with
perpendicular arms, as well as the observing geometry.  The
geometrical dependence manifests itself via the overlap reduction
function (ORF)\cite{Flanagan:1993}, which can be written
as\cite{Whelan:2006}
\begin{equation}
  \label{e:gammadef}
  \gamma_{12}(f)={d_{1ab}}\, {d_2^{cd}}\,
  \frac{5}{4\pi} \iint d^2\Omega_{\nhat}\,
  P^{\text{TT}\nhat}{}^{ab}_{cd}\,
  e^{i2\pi f\nhat\cdot(\vec{r}_2-\vec{r}_1)/c}
\end{equation}
where each IFO's geometry is described by a response tensor
constructed from unit vectors $\hat{x}$ and $\hat{y}$ down the two
arms
\begin{equation}
  d^{ab} = \frac{1}{2}(\hat{x}^a \hat{x}^b - \hat{y}^a \hat{y}^b)
  \ ,
\end{equation}
$\vec{r}_{1,2}$ is the respective interferometer's location and
$P^{\text{TT}\nhat}{}^{ab}_{cd}$ is a projector onto traceless
symmetric tensors transverse to the unit vector $\nhat$.  At zero
frequency, the ORF is determined entirely by detector orientation.
The LHO and LLO sites are aligned as nearly as possible given their
separation on the globe, so that $\gamma_{HL}(0)=-0.89$.
In contrast, the Virgo and GEO600 sites are poorly oriented with
respect to one another, so $\gamma_{GV}(0)=-0.08$.  However, the
frequency-dependence of the ORFs means that the situation is quite
different at frequencies above 40\,Hz, where the IFOs are sensitive.
In particular, the amplitude of $\gamma_{GV}(f)$ does not drop
appreciably for $f$ below about $350\un{Hz}$.
For the other pairs, the behaviour is determined by the
high-frequency limiting form of the ORF,
\begin{equation}
  \label{e:gammalim}
  \gamma_{12}(f)\longrightarrow
  5
  {d_{1ab}}\,P^{\text{TT}\shat_{12}}{}^{ab}_{cd}\,{d_2^{cd}}\,
  \sinc(2\pi f T_{12})
  = \frac{\gamma^{\text{env}}_{12}}{f} \sin(2\pi f T_{12})
\end{equation}
where $T_{12}$ is the light travel time between the detector sites and
$\shat_{12}$ is a unit vector pointing from one site to the other.
While trans-Atlantic light travel times like $T_{\text{HV}}$ and
$T_{\text{LV}}$ are greater than $T_{\text{HL}}$, leading to more
oscillations in the ORF, the overall envelope
$
\gamma_{12}^{\text{env}} =
(5{d_{1ab}}P^{\text{TT}\shat_{12}}{}^{ab}_{cd}{d_2^{cd}}) / (2\pi T_{12})
$
includes geometric projection factors, which more than make up for
this discrepancy, as summarised in Table~\ref{tab:overlap}.
The result is that in the full overlap reduction function
(Fig.~\ref{fig:overlap}) the typical amplitudes
$\abs{\gamma_{\text{HV}}(f)}$ and $\abs{\gamma_{\text{LV}}(f)}$ are
larger than the typical $\abs{\gamma_{\text{HL}}(f)}$ for
$f\gtrsim 200\un{Hz}$.

\subsection{Definition of Sensitivity}

The strength of an isotropic stochastic background can be written in
terms of the one-sided power spectral density $S_{\text{gw}}(f)$ it
would generate in an IFO with perpendicular arms.

\begin{figure}
  \centering
  \includegraphics[width=2.5in,angle=0]{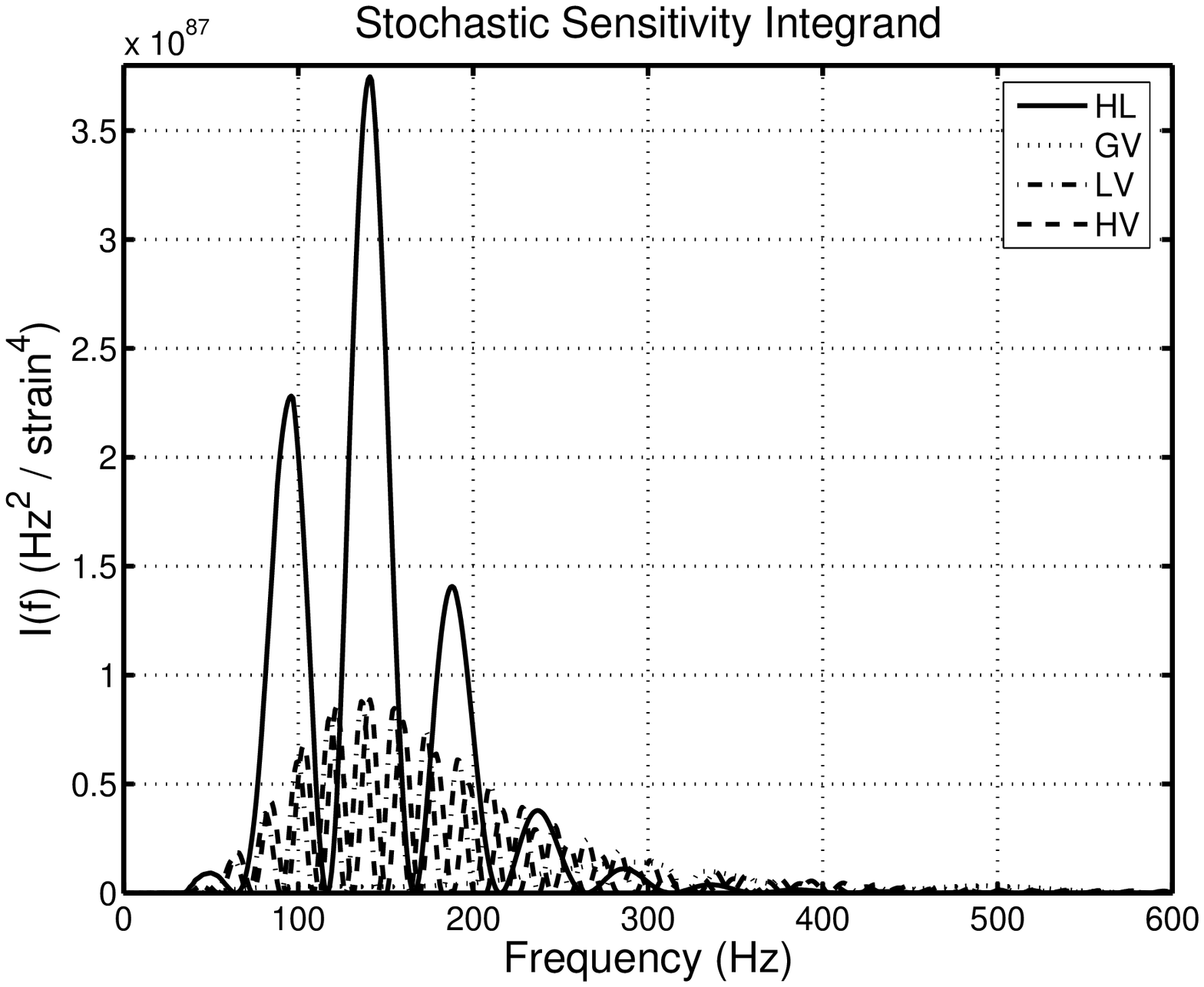}
  \includegraphics[width=2.5in,angle=0]{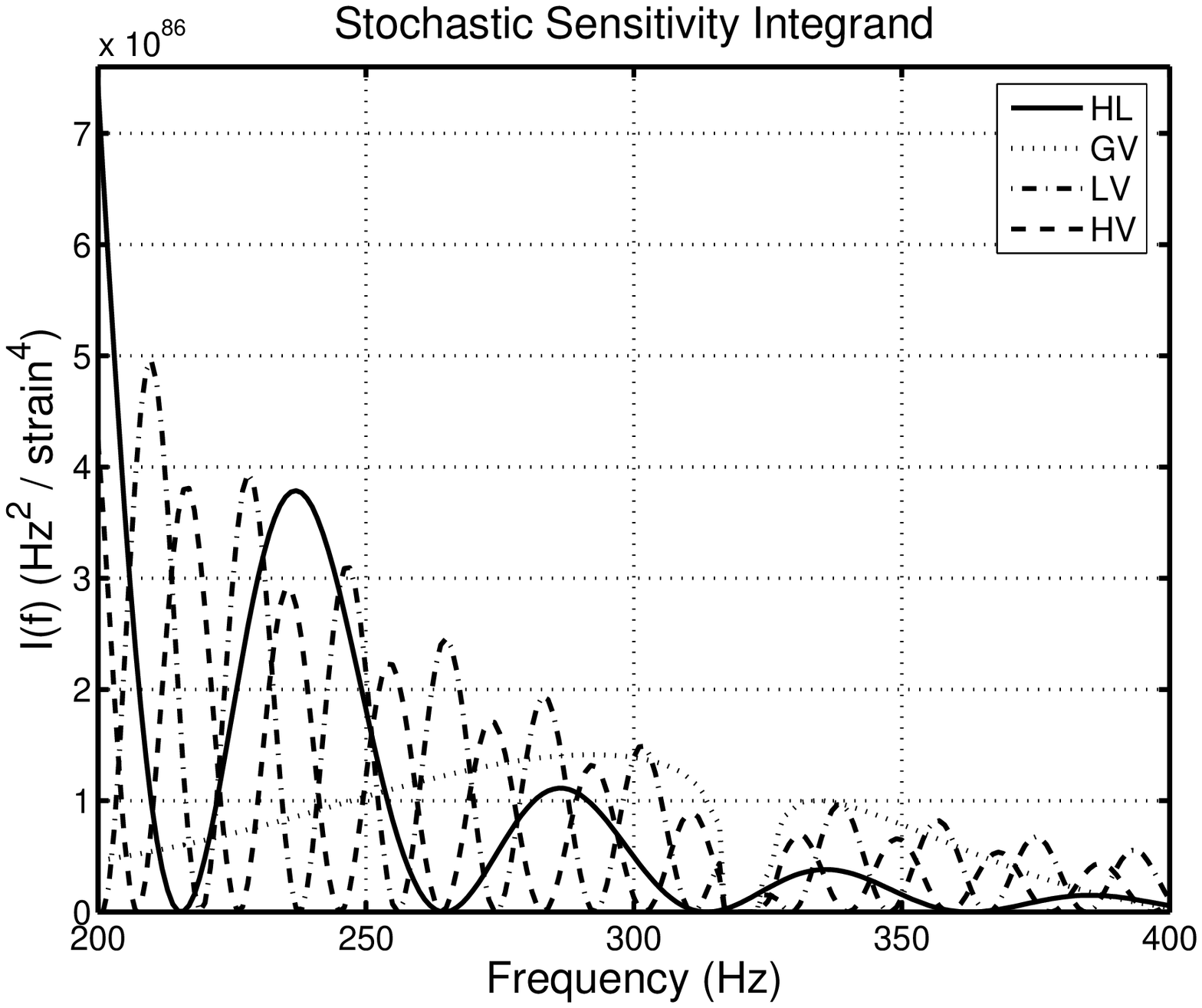}
  \caption{Sensitivity integrands $\mc{I}(f)$ for pairs of detector sites,
    where pairs like HL include e.g., the combined sensitivity of
    H1-L1 and H2-L1.  To get an overall sensitivity, these
    need to be multiplied by the square of the shape
    of the SGWB spectrum $S_{\text{gw}}(f)$.    
    The HL pair is the
    most sensitive for flat and low-frequency spectra, but as shown in
    the closeup at the right, above 200\,Hz, pairs involving Virgo are
    more sensitive when all detectors are operating at their design
    sensitivities.    
    The GH and GL pairs are not shown, since the GEO600 noise spectrum
    \cite{geonoise} means that only the GV pair contributes
    significantly to the overall sensitivity at these frequencies.}
  \label{fig:sensints}
\end{figure}

The standard cross-correlation method seeks to measure the amplitude
$S_R=S_{\text{gw}}(f)/\mc{S}(f)$ of a background whose
$S_{\text{gw}}(f)$ is assumed to have a specified shape $\mc{S}(f)$.
Given co\"{\i}ncident data between times $t_1$ and $t_2$ from
detectors with one-sided noise power spectral densities (PSDs)
$P_{1,2}(f)$, we can make an optimally-filtered cross-correlation
statistic
\begin{equation}
  \label{e:CC}
  Y = \int_{t_1}^{t_2} \int_{t_1}^{t_2} s_1(t)\,Q(t-t')\,s_2(t)\,dt\,dt'
\end{equation}
with the optimal filter defined by its Fourier transform
\begin{equation}
  \widetilde{Q}(f) = \mc{N}\frac{\gamma_{12}(f)\mc{S}(f)}{P_1(f)P_2(f)}
\end{equation}
and $\mc{N}$ chosen so that $\langle Y\rangle=S_R$.  If the geometric
mean of the noise PSDs is large compared to $S_{\text{gw}}(f)$, the
expected variance of the statistic will be
\begin{equation}
  \sigma^2 = \frac{1}{2T} 
  \left(
    \int_{0}^{\infty}
    \frac{\left[\gamma_{12}(f)\,\mc{S}(f)\right]^2}{P_1(f)\,P_2(f)}
  \right)^{-1}
\end{equation}
where $T=t_2-t_1$ is the duration of the data analysed.
The squared signal-to-noise ratio of the standard
cross-correlation statistic will thus be
\begin{equation}
  \mathrm{SNR}^2 := \frac{\langle Y\rangle^2}{\sigma^2}
  = 2T\,S_R^2
  \int_{0}^{\infty}
  [\mc{S}(f)]^2
  \mc{I}_{12}(f)
  \ df
  \ ,
\end{equation}
where we have defined a ``sensitivity integrand'' which illustrates
the contribution to the sensitivity of different frequencies:
\begin{equation}
  \mc{I}_{12}(f) =
  \frac{[\gamma_{12}(f)]^2}{P_{1}(f)\,P_{2}(f)}
\end{equation}
We plot $\mc{I}(f)$ for several pairs of detectors in
Figure~\ref{fig:sensints}, using the design sensitivities
in\cite{ligonoise,virgonoise,geonoise}.
As shown in \cite{Allen:1999}, the optimal method for combining
correlation measurements from different detector pairs is the same as
that for combining measurements from different times: average the
point estimates $Y$ with a relative weighting of $\sigma^{-2}$, and
the resulting variance will be the inverse of the sum of the
$\sigma^{-2}$ values.  This produces a
sensitivity integrand which is the sum of the integrands for
individual pairs:
\begin{equation}
  \mc{I}(f) = \sum_{\text{pair}} \mc{I}_{\text{pair}}(f)
  \ .
\end{equation}
An immediate application of this is to define sensitivity integrands
that combine pairs involving H1 and H2, e.g.,
$\mc{I}_{\text{HL}}=\mc{I}_{\text{H1,L1}}+\mc{I}_{\text{H2,L1}}$.
This is the same as using the spectrum of an optimally combined H
pseudo-detector as described in \cite{Lazzarini:2004}.

\begin{figure}
  \centering
  \includegraphics[width=2.5in,angle=0]{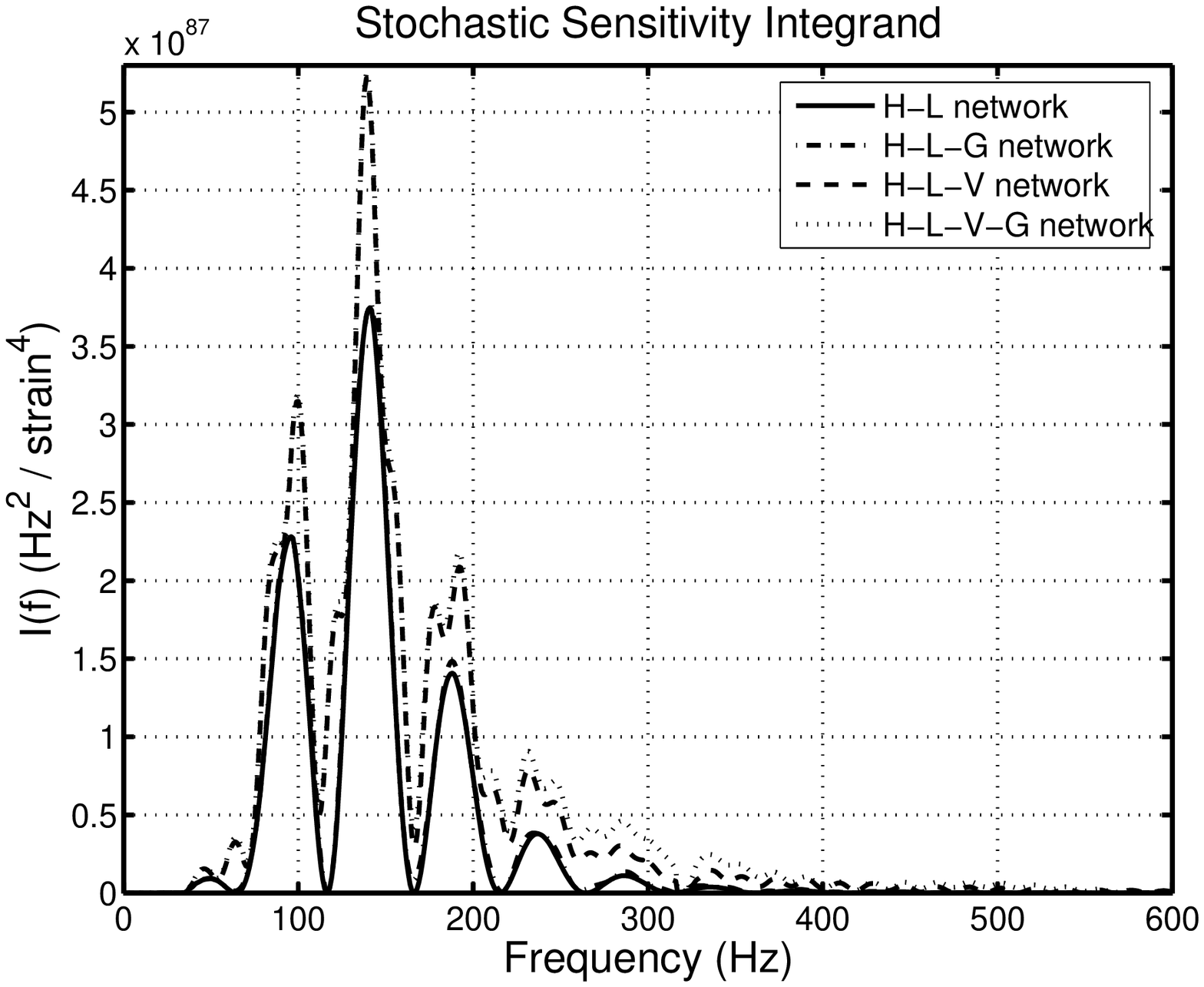}
  \includegraphics[width=2.5in,angle=0]{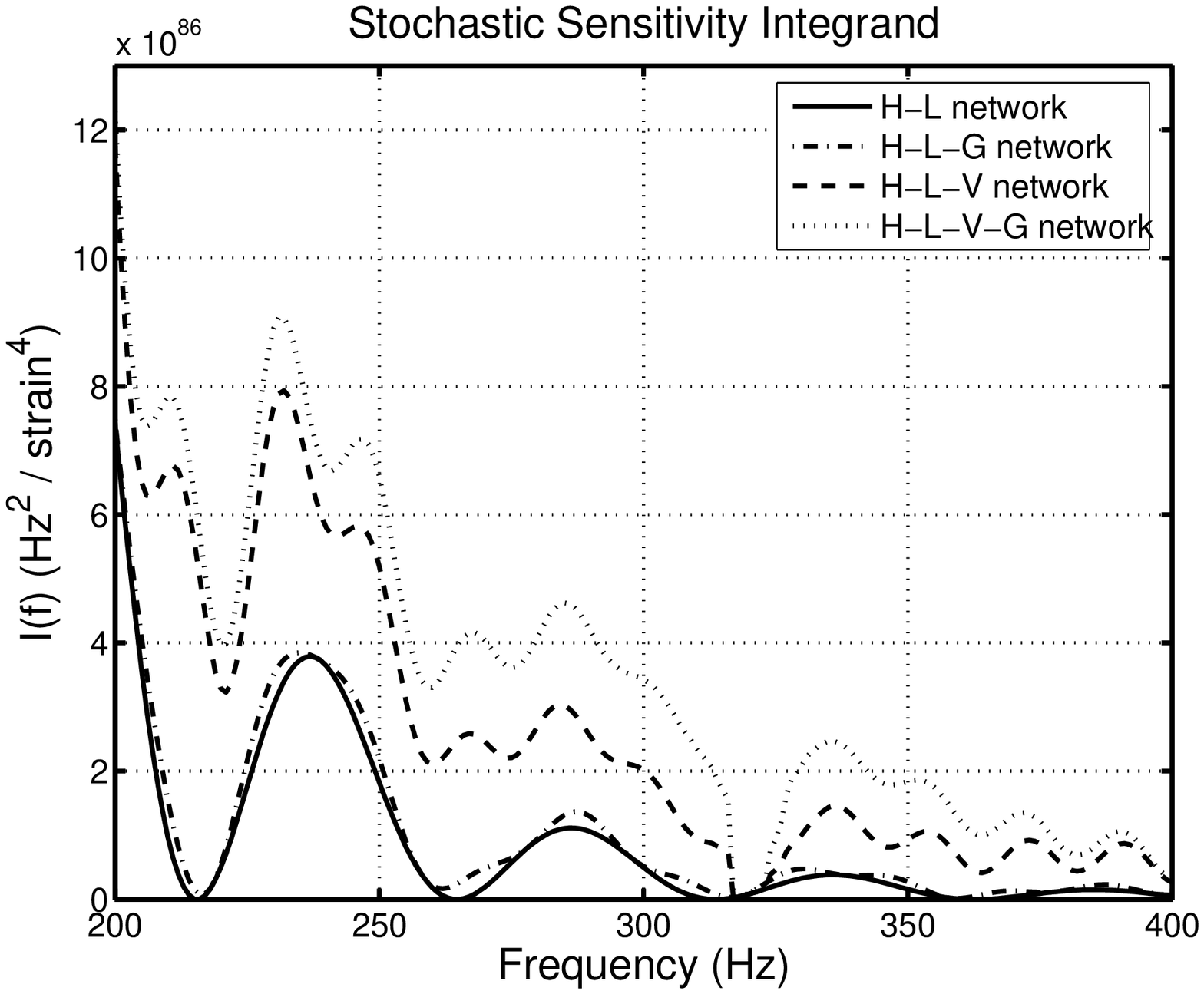}
  \caption{Combined sensitivity integrands for networks of detectors.
    In each case a network including ``H'' includes correlations between
    both H1 and H2 and the detectors at other sites.  As
    the closeup on the right shows, addition of the LV, HV and GV pairs
    to the HL network
    increases sensitivity to backgrounds with significant power above
    200\,Hz.}
  \label{fig:net_sensints}
\end{figure}

Figure~\ref{fig:net_sensints} shows the combined sensitivity for four
networks of detectors operating at design sensitivity: the existing
H-L network, an H-L-G network in which GEO600 is also operating at
design sensitivity, and H-L-V and H-L-V-G networks which also include
a design-sensitivity Virgo.  The H1-H2 pair is not included in
these networks, because the presence of correlated environmental noise
necessitates special treatment of this pair \cite{Fotopoulos:2006}.

Since the power in the faintest detectable ``white'' stochastic
background is proportional to the square root of the area under the
sensitivity integrand, we see that the addition of Virgo to the
LHO-LLO network would be most useful in improving sensitivity to a
narrow-band background peaked above 200\,Hz, or to one whose spectrum
rises with increasing frequency.  As an illustration,
Table~\ref{tab:network_sens} shows for several detector networks the
faintest detectable background with constant $S_{\text{gw}}(f)$ in a
100\,Hz band, assuming one year of observation time and an SNR
threshold of 3.29, associated with 5\% false alarm and false dismissal
rates.
\begin{table}
  \centering
  \caption{Smallest detectable band-limited background using each of
    the detector networks defined in Fig.~\ref{fig:net_sensints}.  In
    each case, this is the strain power spectrum, in units of
    $10^{-48}\un{Hz}^{-1}$, that could be detected with 5\% false alarm
    and 5\% false dismissal rates, using one year of co\"{\i}ncident
    data at design sensitivity.}
  \input{network_sens}
  \label{tab:network_sens}
\end{table}

\section{Simulations}

\label{s:simulations}

To test cross-correlation analyses of LIGO and Virgo data, we injected
a simulated SGWB signal into simulated LIGO and Virgo noise.  The
simulated noise data were the 24 hours of H1, H2, L1, and V1 data, all
at nominal design sensitivity, initially generated for simulated
searches for GW bursts and inspiralling compact object binaries
\cite{LVBurst,LVInspiral}, known as ``project 1b''.  We chose a
spectral shape designed to highlight the performance of the LV and
HV pairs at $f\gtrsim200\un{Hz}$, but which corresponds to a model of
an astrophysical SGWB.  The spectrum we used
is associated with the superposition of the tri-axial
emission from the extra-galactic population of spinning magnetars with
type I superconducting interior, as described in model B of
\cite{Regimbau:2006}, but updated using the star formation history of
\cite{Hopkins:2006}.

\begin{figure}
  \centering
  \includegraphics[width=2.5in,angle=0]{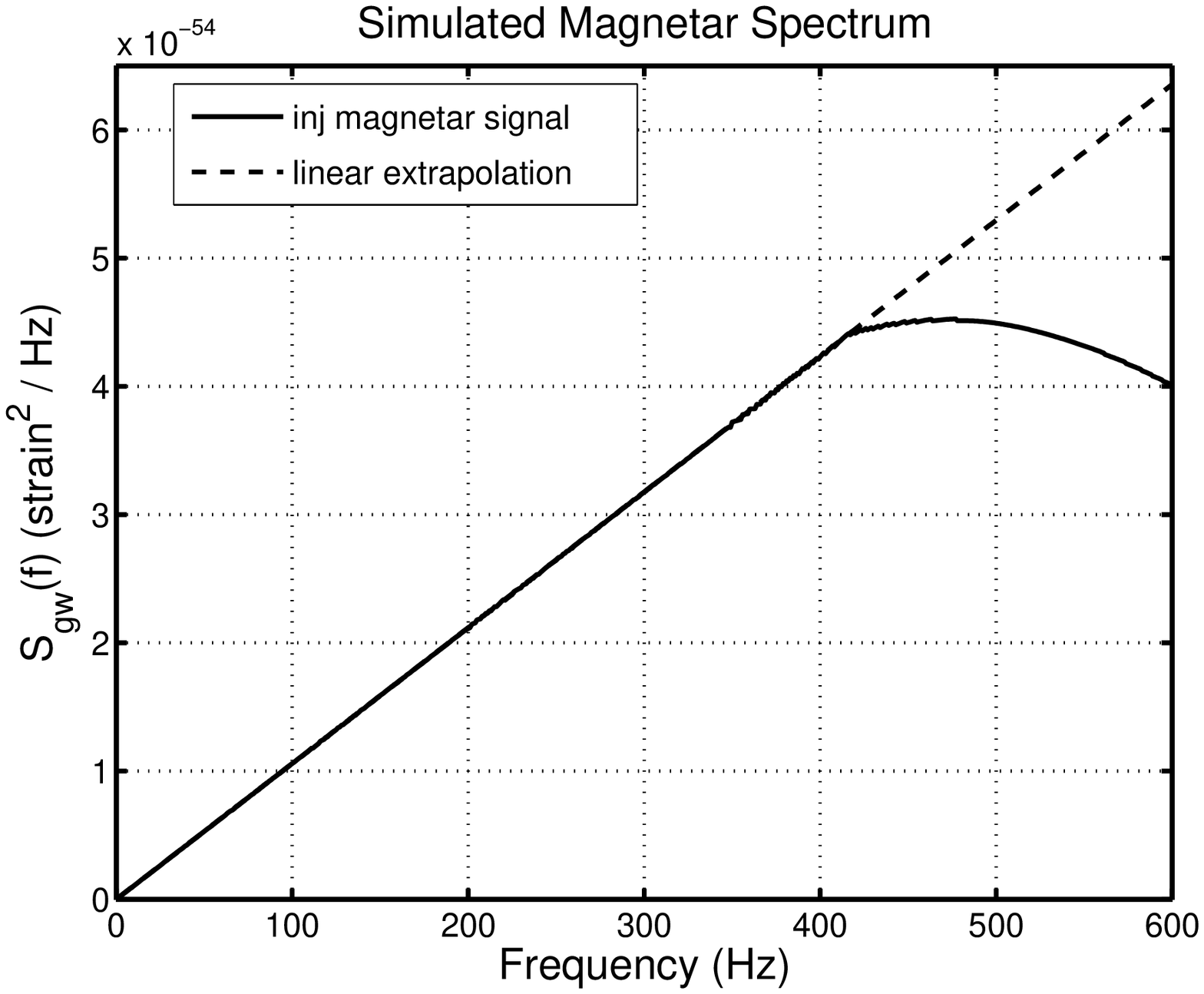}
  \includegraphics[width=2.5in,angle=0]{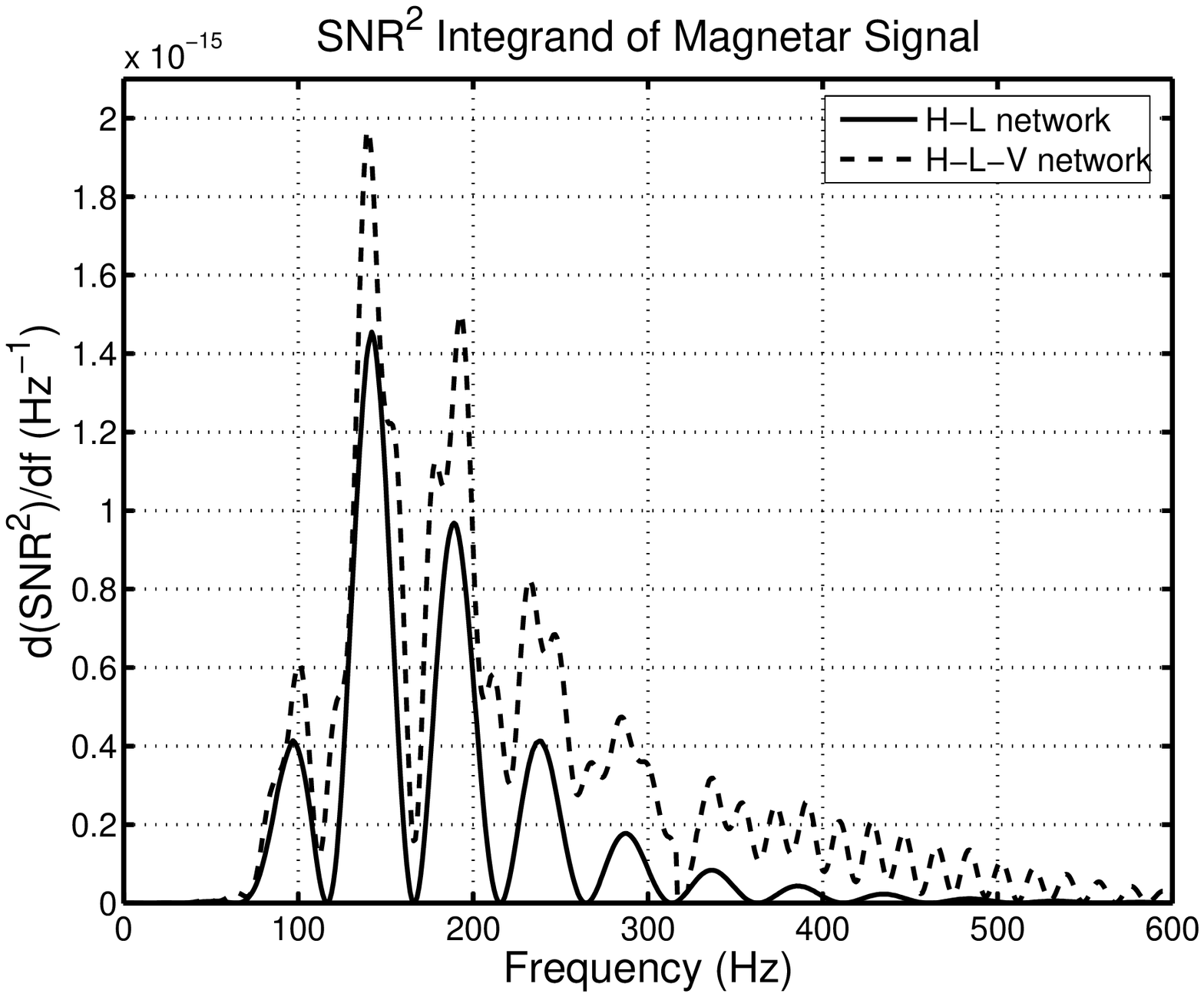}
  \caption{The magnetar spectrum used to generate simulated signals,
    and the associated integrand for squared signal-to-noise-ratio.
    Note that since $S_{\text{gw}}(f)$ increases approximately
    linearly with frequency up to about 400\,Hz, this spectrum, while
    broad-band, tends to favour the higher frequencies where LIGO-Virgo
    detector pairs are more sensitive.  Beyond 400\,Hz, the spectrum
    is no longer linear, but we show the SNR that would result from
    attempting to detect it with a $\mc{S}(f)\propto f$ filter; the
    corrections are negligible below about 500\,Hz, and still small
    throughout the frequency range displayed.  Note that integrating
    the area under the H-L-V curve on the right still gives an SNR
    below $10^{-6}$ from 24 hours of data, so we scale up the injected
    strain signal by a factor of several thousand in the simulations
    described in Section~\protect\ref{s:simresults}.}
  \label{fig:magnetar}
\end{figure}

In Figure~\ref{fig:magnetar} we show the
spectrum and the associated sensitivity integrand in the corresponding
detectors.
Since this spectrum rises with increasing frequency up to about
400\,Hz, it is useful for illustrating the utility of a network
involving Virgo to search for a broad-band astrophysical source.
Since the model signal would be too weak to detect with
first-generation interferometers, we scale up the signal strength,
injecting a signal with the same spectral shape, but a much
larger amplitude, in our simulations.

\subsection{Simulation algorithm}

The problem of simulation of the signal in a pair of detectors due to
an isotropic and Gaussian SGWB has been considered previously in e.g.,
\cite{Allen:1999,Bose:2003,LLOALLEGRO}.  For this work we generalise
that to a network of $N$ GW detectors.\footnote{In our case, to
  simulate signals in H1, H2, L1, and V1, $N=3$, because H1 and H2
  have the same response tensor and therefore the same simulated GW
  signal can be used for both of them.}  We need to satisfy
\eqref{eq:h1h2corr} for each pair of detectors; treating $\{h_A(f)\}$
as the elements of a column vector $\mathbf{\widetilde{h}}(f)$ and
$\{\gamma_{AB}(f)\}$ as the elements of a real, symmetric
matrix\footnote{For non-isotropic backgrounds, the ORF is complex
  rather than real, and more care must be taken with the definition of
  the Hermitian matrix $\boldsymbol{\gamma}(f)$.}
$\boldsymbol{\gamma}(f)$, we can write this as a matrix equation
\begin{equation}
  \label{eq:covariance}
  \langle \mathbf{\widetilde{h}}(f)\,\mathbf{\widetilde{h}}(f')^\dagger \rangle
  = \frac{1}{2}S_{\text{gw}}(f) \boldsymbol{\gamma}(f) \delta(f-f')
\end{equation}
If we can define a matrix $\boldsymbol{\beta}(f)$ which factors
$\boldsymbol{\gamma}(f)$:
\begin{equation}
  \label{eq:factorise}
  \boldsymbol{\gamma}(f)=\boldsymbol{\beta}(f)\boldsymbol{\beta}(f)^\dagger
  \ ,
\end{equation}
then we can generate $N$ independent white noise data streams
$\{\widetilde{\eta}_A(f)\}$ which satisfy
\begin{equation}
  \langle \boldsymbol{\widetilde{\eta}}(f)
  \,\boldsymbol{\widetilde{\eta}}(f')^\dagger \rangle
  =  \boldsymbol{1} \delta(f-f')
\end{equation}
and then convert them into the desired coloured correlated data streams
via
\begin{equation}
  \label{eq:fdsim}
  \mathbf{\widetilde{h}}(f) = \sqrt{\frac{S_{\text{gw}}(f)}{2}}
  \boldsymbol{\beta}(f)\boldsymbol{\widetilde{\eta}}(f)
\end{equation}
For a given $\boldsymbol{\gamma}(f)$, there are different choices of
$\boldsymbol{\widetilde{\eta}}(f)$ which achieve the factorisation
\eqref{eq:factorise}.

Since \eqref{eq:covariance} is a covariance matrix, it is positive
semi-definite, from which it follows (since $S_{\text{gw}}(f)>0$) that
$\boldsymbol{\gamma}(f)$ is positive semidefinite as well.
$\boldsymbol{\gamma}(f)$ could have one or more zero eigenvalues in
the presence of linear dependence between detector outputs.  A
practical example of this is two detectors sharing the same geometry
and location, such as H1 and H2, are included in the
network.\footnote{Less trivial examples can be constructed, for
  example three detectors in the same location and in the same plane.}
We avoid this problem by generating a simulated signal for H1 and
injecting it into both H1 and H2.

In the generic case where $\boldsymbol{\gamma}(f)$ is positive
definite, we can make the straightforward choice of the Cholesky
decomposition\cite{Golub:1989}, in which $\boldsymbol{\beta}(f)$ is a
lower diagonal matrix.  In the case $N=2$, this reduces to the form
used in e.g.,\cite{Allen:1999}.  For $N=3$, if the diagonal elements
of $\boldsymbol{\gamma}(f)$ are unity\footnote{This is the case for
  interferometers with perpendicular arms, but \emph{not} for GEO600
  or for resonant bar detectors; see \cite{Whelan:2006}}, the explicit
form is
\begin{equation}
  \boldsymbol{\beta}(f) =
  \left(\begin{array}{ccc}
      1 & 0 & 0\\
      \gamma_{12} & \sqrt{1-\gamma_{12}^{2}} & 0 \\
      \gamma_{13} 
      & \frac{\gamma_{23}-\gamma_{12}\gamma_{13}}{\sqrt{1-\gamma_{12}^{2}}}
      & \sqrt{\frac{1+2\gamma_{12}\gamma_{13}\gamma_{23}-\gamma_{12}^{2}-\gamma_{13}^{2}-\gamma_{23}^{2}}{1-\gamma_{12}^{2}}}\end{array}\right)
\end{equation}
However in practice we can simply use a fast iterative algorithm for
the Cholesky decomposition.

Other factorisation strategies which treat the different detectors
more symmetrically (e.g., defining
$\boldsymbol{\beta}(f)=\boldsymbol{\Lambda}(f)^{1/2}\mathbf{U}(f)$
where $\boldsymbol{\Lambda}(f)$ is the diagonal matrix of eigenvalues
of $\boldsymbol{\gamma}(f)$ and $\mathbf{U}(f)$ is the matrix
constructed from the corresponding eigenvectors) may be more
demanding in terms of computational power, but more numerically stable
when correlations between the detectors are large and off-diagonal
elements of $\boldsymbol{\gamma}(f)$ are comparable to unity. In
particular, this strategy can deal directly with the case when
$\boldsymbol{\gamma}(f)$ has one or more zero eigenvalues.

\subsection{Filtering strategy}

The continuous frequency-domain idealisation \eqref{eq:fdsim} needs to
be applied with care to finite stretches of real detector data.  In
the time domain, the multiplication \eqref{eq:fdsim} amounts to a
convolution
\begin{equation}
  \mathbf{h}(t) = \int_{-\infty}^{\infty}
  \mathbf{K}(t-t')\,\boldsymbol{\eta}(t')\,dt'
\end{equation}
with a kernel which is the inverse Fourier transform of
\begin{equation}
  \mathbf{\widetilde{K}}(f) =
  \sqrt{\frac{S_{\text{gw}}(f)}{2}} \boldsymbol{\beta}(f)
  \ .
\end{equation}
If the time-domain kernel $\mathbf{K}(\tau)$ is negligible outside the
interval $-\tau_{-} < \tau < \tau_{+}$, we can use the standard
overlap-and-add strategy to generate a continuous stream of
time-series data, as illustrated in Figure~\ref{fig:OverlapAdd}:
\begin{figure}
  \includegraphics[width=1\columnwidth]{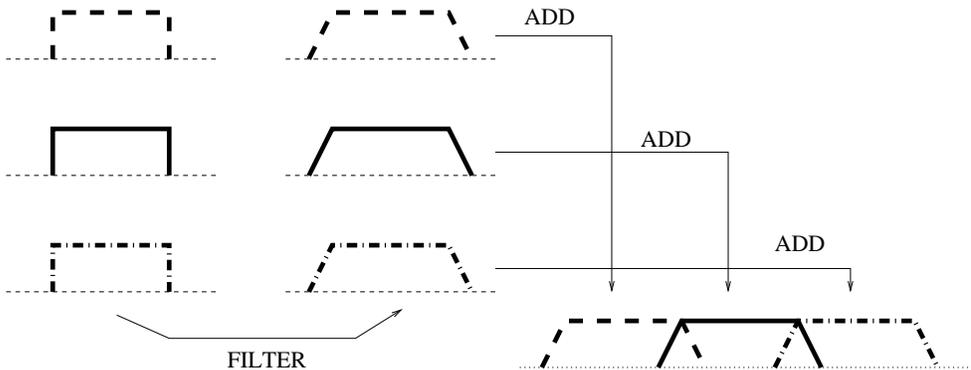}
  \caption{\label{fig:OverlapAdd}The overlap-and-add algorithm.}
\end{figure}
\begin{enumerate}
\item Generate a sequence of ``buffers'' of white noise data for each
  of the $N$ detectors, each of length $T$.
\item Convolve each buffer with the kernel $\mathbf{K}(\tau)$ to
  obtain a time series of length $\tau_{-}+T+\tau_{+}$, with an
  associated start time $\tau_{-}$ before the start and end time
  $\tau_{+}$ after the end of the original buffer.  (This is most
  naturally done in the frequency domain, zero-padding the white noise
  by $\tau_{-}$ at the beginning and $\tau_{+}$ at the end, then
  Fourier transforming and multiplying by $\mathbf{\widetilde{K}}(f)$
  before inverse-Fourier-transforming.)
\item Add together the processed data buffers, overlapping by
  $\tau_{+}$ on one end and $\tau_{-}$ on the next, producing
  correlated coloured time-series data of duration $T$ times the number
  of buffers, plus transients of $\tau_{-}$ at the beginning and
  $\tau_{+}$ at the end, which are discarded.
\end{enumerate}

This strategy was implemented in code based on Virgo's ``Noise
Analysis Package'' ({\sc nap})~\cite{NAP}.

\section{Analysis of simulated data}
\label{s:simresults}

\begin{figure}
  \begin{center}
     \includegraphics[width=3in,angle=0]{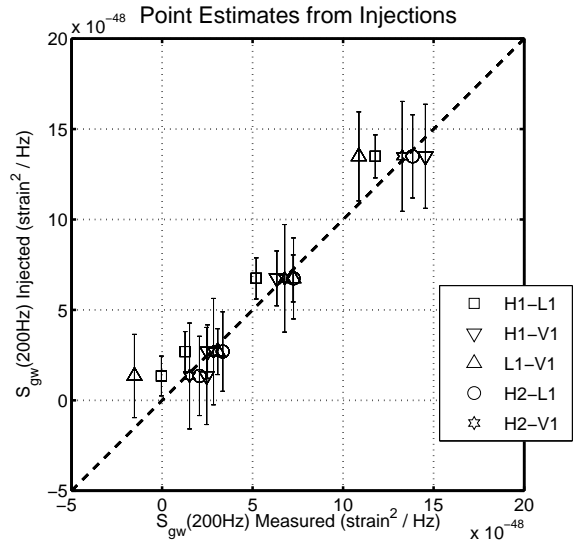}
  \end{center}  
  \caption{The individual point estimates and one-sigma error bars
    with signals injected at two different levels.  The strength of
    the injection is shown on the vertical axis, as a strain power at
    200\,Hz, while the estimate of this quantity from each detector
    pair is on the horizontal axis.  The error bars are drawn
    vertically for ease of reading, and because they are frequentist
    error bars as quoted.  Note that the signals are seen at
    comparable strength in all five pairs of detectors.}
  \label{fig:p1b}
\end{figure}

The continuous signals described in Section~\ref{s:simulations} were injected
into the ``project 1b'' simulated noise and the resulting time series
analysed using the {\sc matapps} stochastic analysis code developed by the
LSC. \cite{matapps}.
In particular, the cross-correlation \eqref{e:CC} was performed in the
frequency domain without any need to resample the time-series data,
using a variation of the method described in \cite{Whelan:2005} and
applied in \cite{LLOALLEGRO}.
Several simulation runs were performed in which
same set of simulated signals were injected into the four data
streams, scaled up by a different factor for each run.  The results of
two of those simulations are shown here.  In Figure~\ref{fig:p1b} we
plot the individual point estimates and error bars in each of the five
detector pairs (every combination except for H1-H2).  The amplitude
measure $S_R$ quoted is the one-sided strain PSD at 200\,Hz.
The optimal filter used assumed the shape $\mc{S}(f)\propto f$,
although the injected signals only had that form below about 400\,Hz.
The analysis was done over a frequency band 50--500\,Hz to avoid
difficulties arising from this mismatch, as illustrated in
Figure~\ref{fig:magnetar}.

The optimal combination of these results is shown in
Table~\ref{tab:p1b}, both for the full network and for the network
consisting only of the two LHO-LLO pairs.  Our error bars are reduced
by 15\% via the inclusion of the LIGO-Virgo pairs.  Note that although
this source has more power at the intermediate frequencies favoured by
the LIGO-Virgo pairs, it is still broad-band.  A narrow-banded source
with most of its power above 200\,Hz would favour LIGO-Virgo pairs to
an even greater extent.
\begin{table}
  \input{project1b_results}
  \centering
  \caption{The values of $S_{\text{gw}}(200\un{Hz})$ calculated from
    the simulated project 1b data, with associated one-sigma error bars,
    for the H-L network consisting of the H1-L1 and H2-L1 pairs, and
    for the H-L-V network consisting of those plus the H1-V1, H2-V1,
    and L1-V1 pairs.  We see that including Virgo in the network reduces
    our error bars for this broad-band astrophysical spectrum by 15\%.}
  \label{tab:p1b}
\end{table}

\section{Conclusions and Outlook}

We have demonstrated how the inclusion of LIGO-Virgo and possibly
GEO600-Virgo detector pairs can enhance the sensitivity of the global
GW detector network to an isotropic background of gravitational waves,
particularly at frequencies above 200\,Hz.  As a practical
illustration, we have adapted and applied pipelines for generating
correlated simulated signals in the LSC and Virgo detectors, and for
analysing co\"{\i}ncident data via the standard cross-correlation
technique.  The specific astrophysical model we used
(which was chosen because its frequency spectrum was peaked at
frequencies where LIGO-Virgo pairs at design will be more sensitive
than LLO-LHO pairs)
had to have its
amplitude increased to be detectable by any pair of first-generation
IFOs.  Nonetheless, the exercise illustrates
how multiple detector pairs can be
used to discover an ``unexpected'' background.

Virgo is not yet at its nominal design sensitivity, but has improved
its sensitivity markedly over the past year, and its
first full science run starts in May 2007, to be analysed
in conjunction with the end of LIGO's S5 run.


\ack

The authors would like to thank their colleagues in the LIGO
Scientific Collaboration and the Virgo project.  JTW gratefully
acknowledges Loyola University New Orleans and the University of Texas
at Brownsville.
This work was supported by the National Science Foundation under grant
PHY-0300609 and by the Max-Planck-Society.
This paper has been assigned LIGO Document Number
{\dcc} and AEI document number AEI-2007-017.

\end{document}

%% file: gammaDC.tex
  \begin{tabular}{@{}ccccccc}
& HL & HV & LV & GH & GL & GV \\
    \mr
$\gamma_{12}(0)$ &-0.89 & -0.02 & -0.25 & 0.42 & -0.32 & -0.08\\
    $(2\pi T_{12})^{-1}$ (Hz) & 15.9 & 5.8 & 6.0 & 6.3 & 6.3 & 49.8 \\ 
    $\gamma_{12}^{\text{env}}$ (Hz) &  -1.7 & 5.0 & -5.9 & 3.0 & 4.3 & -17.2 \\ 
  \end{tabular}

%% file: network_sens.tex
  \begin{tabular}{@{}ccccc}
    \\
 Band & H-L & H-L-G & H-L-V & H-L-V-G
    \\
    \mr
  200--300\,Hz & 5.79 & 5.43 & 3.44 & 3.04
    \\
  300--400\,Hz & 18.57 & 15.37 & 7.92 & 5.88
  \end{tabular}

%% file: project1b_results.tex
  \begin{tabular}{@{}lll}
    \\
\multicolumn{3}{c}{$S_{\text{gw}}(200\un{Hz})$ ($10^{-48}\un{Hz}^{-1}$)}    \\
 Injected & H-L Result & H-L-V Result \\    \mr
$1.35$& $0.39\pm0.98$& $0.43\pm0.82$    \\
$2.70$& $1.69\pm0.99$& $2.31\pm0.67$    \\
$6.74$& $5.63\pm1.02$& $6.29\pm0.69$    \\
$13.49$& $12.22\pm1.06$& $12.35\pm0.88$    \\
  \end{tabular}